\documentclass[pra,preprint,showpacs]{revtex4}

\usepackage{graphicx}

\newcommand\vect[1]{\mathbf{#1}}

\begin{document}

\title{Local-field excitations in 2D lattices of resonant atoms}

\author{S.~N.~Volkov}
\author{A.~E.~Kaplan}
\affiliation{Dept.\ of Electrical and Computer Engineering, 
Johns Hopkins University, Baltimore, MD 21218}

\date{February 12, 2010}

\begin{abstract}

We study excitations of the local field (locsitons) in nanoscale 
two-dimensional (2D) lattices 
of strongly interacting resonant atoms and various unusual effects associated 
with them.
Locsitons in low-dimensional systems and the resulting spatial strata and more 
complex patterns on a scale of just a few atoms were predicted by us earlier 
[A.~E.\ Kaplan and S.~N.\ Volkov,
Phys.\ Rev.\ Lett.\ \textbf{101}, 133902 (2008)].
These effects present a radical departure from the classical Lorentz-Lorenz 
theory of the local field (LF), which assumes that the LF is virtually uniform 
on this scale.
We demonstrate that the strata and patterns in the 2D lattices may be 
described as an interference of plane-wave locsitons,
build an analytic model for such unbounded locsitons, and derive and analyze 
dispersion relations for the locsitons in an equilateral triangular lattice.
We draw useful analogies between one-dimensional
and 2D locsitons, but also show that the 
2D case enables locsitons with the most diverse and unusual properties.
Using the nearest-neighbor approximation, we find the locsiton 
frequency band for different mutual orientations of the lattice and the 
incident field.
We demonstrate a formation of distinct vector locsiton patterns consisting of 
multiple vortices in the LF distribution and suggest a way to design finite 2D 
lattices that exhibit such patterns at certain frequencies.
We illustrate the role of lattice defects in supporting localized locsitons
and also demonstrate the existence of ``magic shapes'', for which
the LF suppression at the exact atomic resonance is cancelled.

\end{abstract}


\preprint{Submitted to \emph{Phys.~Rev.~A}}

\maketitle
\label{sec-intro}

\section{Introduction}

In our recent Letter \cite{Kaplan2008PRL} we predicted nanoscale field 
patterns (\emph{stratification}) emerging in one-dimensional (1D) arrays and 
two-dimensional (2D) lattices of strongly interacting atoms, driven by a 
radiation nearly resonant to the atomic transition.
We predicted excitation of so called \emph{locsitons} and a host of 
related effects.
A general formulation of the problem and a more detailed theory for 1D arrays 
was presented in our most recent paper \cite{Kaplan2009PRA}.
The present paper is an extension of \cite{Kaplan2008PRL,Kaplan2009PRA} toward 
the theory of \emph{2D lattices} of resonant atoms, which produce a much 
richer set of effects.
We construct here a detailed theory of interactions in the system by developing 
different 2D versions of the nearest-neighbor approximation (NNA), 
including the ``near-ring'' approximation (NRA).
We also derive dispersion relations for various lattice-polarization 
configurations for all locsiton wave vectors within the corresponding first 
Brillouin zones.
Our theory predicts such phenomena as subwavelength multicell patterns, 
including multivortex locsiton excitations, and locsitons localized near
lattice defects.
Further on, we predict ``magic shapes'' of nanosize groups of atoms, which 
reverse the effect of a resonant locsiton suppression 
present in all but few configurations.
The simplest magic configuration which can be cut out of a triangular lattice 
is a six-point star with an atom at its center, which makes the lowest 
``magic number'' of atoms to be 13.

The predicted effects would be totally unexpected within the standard theory 
of local fields \cite{BornWolf} going back to the works of 
Lorentz \cite{Lorentz1880} and Lorenz \cite{Lorenz1881}.
That celebrated theory asserts that the microscopic electric field 
$\vect{E}_\mathrm{L}$ acting upon any given atom in a medium---the 
\emph{local} field (LF)---differs from the macroscopic field $\vect{E}$ of the 
electromagnetic wave, because electric dipoles induced in neighboring atoms 
produce extra field to supplement the field of the incident wave.
This difference is significant in dense media, where the interatomic 
interactions are sufficiently strong.
Under such conditions, typical interatomic distances are much shorter than the 
optical wavelength, and the dipole-dipole interactions between atoms can be 
treated as quasistatic.
The major point of the Lorentz-Lorenz theory (LLT) of local fields 
is that the LLT contains a fundamental assumption 
(which often remains implicit and unspoken in the literature)
that the LF varies very insignificantly 
between neighboring atoms, 
much like the applied optical field on the subwavelength scale.
Unsurprisingly, that theory results in the LF being proportional to the 
macroscopic electric field, 
$\vect{E}_\mathrm{L} = \vect{E} (\epsilon + 2)/3$, where 
$\epsilon$ is the dielectric constant of the medium.

As we have shown in \cite{Kaplan2008PRL,Kaplan2009PRA}, this assumption of the 
LF uniformity is not universally applicable; moreover, it
completely falls apart when the uniformity
of the atomic lattice is disturbed by impurities, boundaries, etc.
Indeed, when the interatomic interactions are 
sufficiently strong and the system 
is not very large, highly nonuniform 
LF distributions emerge, resulting in a 
strong stratification of the LF and atomic excitations.
This effect is best manifested in small-scale ordered arrays and lattices of 
atoms at near-resonance conditions, which allow to attain high interaction 
strength between neighboring atoms and to easily control it by tuning the 
laser frequency.
We have shown \cite{Kaplan2008PRL,Kaplan2009PRA} that when the interaction 
with neighboring atoms becomes comparable to that with the external field, so 
that the interaction strength exceeds some critical value, 
the system will support LF excitations, which we call \emph{locsitons}.
In finite-size arrays and lattices, standing waves of locsitons may form 
nanoscale strata and complex patterns in the LF (and hence, in the induced 
atomic dipoles).
A typical spacing between atoms in the arrays and lattices exhibiting the LF 
stratification is a few orders of magnitude shorter than the wavelength of 
light, so the quasistatic approximation of the standard LLT
can still be used.

It is worth noting that locsitons are basically a
linear phenomenon and can be excited by a weak incident field. 
We want to stress that locsitons, i.\,e., spatially nonuniform solutions,
are not new stable-state alternatives to
a presumably unstable uniform Lorentz solution at 
certain interaction parameters;
the stability or instability of the solution is not an issue here.
The emergence of locsitons is determined by the 
boundary conditions in a \emph{finite structure}, 
so a locsiton is essentially \emph{the only}
physical solution, which thus replaces the uniform Lorentz solution.
Absorption plays an important role here, as it directly affects spatial 
attenuation of locsitons and thus the maximum distance to a boundary,
defect, or other inhomogeneity where locsitons can appear.
In particular, as we have shown in \cite{Kaplan2009PRA}, the size of an array
that can support well pronounced locsitons is directly
related to the characteristic absorption length.
Locsitons vanish in the bulk of a crystal 
sufficiently far away from boundaries or defects.

Due to recent advances in fabricating nanoscale structures, the
observation and practical applications of the LF nanostratification 
are becoming a reality.
Theoretically, strongly interacting resonant particles discussed in 
\cite{Kaplan2008PRL} do not have to be atoms, but may also be quantum 
dots, molecules, clusters, etc.
However, one has to remember that one
of the major conditions for the structure
to support the locsitons is that the interaction strength 
has to exceed a critical value.
Since this strength is proportional to the square of an individual 
dipole momentum and inversely proportional to
the cube of the interparticle spacing and the linewidth
of the particle resonance (see below), the atoms
may become preferred candidates.
A very high finesse of atomic resonances (i.\,e.,
their narrow linewidth), compared, for example, to plasmons (see below),
also contributes greatly to the phenomenon,
allowing one to see high-order locsiton resonances.

To observe locsitons, one have to create
conditions to couple them efficiently 
to an optical or some other kind of a probe.
In \cite{Kaplan2008PRL,Kaplan2009PRA} we 
suggested a few promising methods of locsiton detection.
In particular, locsitons could be observed 
via size-related resonances in a scattering 
of laser radiation or via x-ray or electron-energy-loss spectroscopy.

There are many potential applications of locsitons;
here we will mention two of them which were discussed in \cite{Kaplan2008PRL}.
It was shown in \cite{Kaplan2008PRL} that
in the presence of a sufficiently strong optical field
(i.\,e., in the nonlinear case),
the LF in 1D arrays of strongly coupled dipoles 
can exhibit optical bistability, 
which could be used to design nanoscale 
all-dielectric logic elements and switches.
Such devices might complement currently used semiconductor-based electronic
circuits.
Another potential application of locsitons could be based on the extreme 
sensitivity of size-related locsiton resonances to the size and shape of the
system.
At the exact atomic resonance, the field is normally ``pushed out'' of the 
atomic system, unless it has a certain ``magic shape'' \cite{Kaplan2008PRL}.
Consequently, such ``magic structures'' of atoms could find applications in 
designing nanoscale biosensors.

Removing the assumption of the LLT that the 
dipoles in the medium oscillate in lockstep with the incident electromagnetic 
wave is a substantial paradigm shift in the 
theory of light-matter interaction.
Locsitons predicted within our broader approach are a new 
phenomenon, although we can provide some incomplete 
but illustrative analogies from other areas of physics.
For example, short- and long-wavelength strata in \cite{Kaplan2008PRL} are 
reminiscent of ferromagnetic and antiferromagnetic arrangements of static 
magnetic dipoles in the Ising model.
The LLT is, on the other hand, more similar 
to the mean-field approach of the Curie-Weiss theory for magnetic media 
\cite{Aharoni-book}.
The Ising model is known to have richer consequences than the Curie-Weiss 
theory.
Our case is, however, substantially different and 
most of all, more versatile than the Ising model. 
Indeed, instead of being static, as in the Ising model,
the atomic dipoles are induced by the 
applied optical field and can oscillate 
with arbitrary amplitude and phase.
By their nature, locsitons may be classified as Frenkel excitons 
\cite{Kittel-book}, because there is no charge transfer between atoms and 
the dipole interaction is due to bound electrons.
Some of the locsiton effects, first of all, wave resonances,
may be viewed as analogues of other types of oscillations
and waves in condensed matter, like plasmons and phonons \cite{Kittel-book},
as well as low-dimensional effects, like surface 
plasmons \cite{Shalaev2005,Markel2007}, size-related resonances in thin metal 
films \cite{Sandomirskii1967} or long organic molecules \cite{Chernyak2001},
``quantum carpets'' \cite{Kaplan2000}, arrays of pendulums
or electronic circuits \cite{Kaplan2009PRA},  etc.
Approaching from another perspective, 
one can view the formation of a locsiton 
band as a Rabi broadening of the atomic resonance 
due to strong interatomic interactions,
which is essentially similar to the band formation in solid-state theory.
However, although all waves and oscillations may be said to have something 
in common, locsitons form a distinct new class of phenomena because of the 
nonconductive, dielectric, nature of their optical response and a strong 
coupling between atoms, which is needed for attaining dramatic size-related 
resonances, magic configurations, and other interesting effects.
A separate issue outside of the scope of this paper
is how the structures can be fabricated or arranged.
One can envision placing the atoms
in a controlled way on the surface of suitable
dielectric materials;
recent developments in the atomic- and ion-traps
technology allow for arranging
atoms in vacuum as 1D arrays and 2D ``crystals''
in the so called wire traps \cite{iontrap}.

Our paper is structured as follows.
In Sec.~\ref{sec-general} we outline our problem and present some general 
formulas;  for more details the reader should refer to our recent paper
\cite{Kaplan2009PRA}.
In Sec.~\ref{sec-in-plane} we describe locsitons in infinite, unbounded, 2D 
lattices of resonant atoms in the case when the incident field is polarized in 
the lattice plane.
The equations for the LF and the dispersion relations for the 
locsitons are first obtained in the NRA and 
then in the more precise NNA.
In Sec.~\ref{sec-finite} we discuss effects that arise in finite 2D 
lattices, in particular, formation of 2D patterns of the LF, due to 
size-related locsiton resonances, and the ``magic'' cancellation of the 
resonant LF suppression.
In Sec.~\ref{sec-normal} we describe 2D locsitons in the case when the 
incident field is normal to the lattice plane.
Sec.~\ref{sec-conclusions} summarizes main results of our paper.

\section{2D lattices:  General model}
\label{sec-general}

Let us consider a 2D lattice of strongly resonant identical particles, which 
we will further call ``atoms''.
We will assume that, for the incident laser frequency $\omega$ near their 
resonant frequency $\omega_0$, these atoms can be described by a two-level 
model with the transition dipole moment $d_a$.
In the linear case, i.\,e., when the laser intensity
is significantly lower than the saturation intensity,
the results for the two-level model coincide exactly with those
for the classical harmonic oscillator, see \cite{Kaplan2009PRA}.
We further only consider lattices of atoms interacting via quasistatic 
near-field dipole forces.
In general, this assumption is valid if, 
on one hand, the minimum separation $l_a$ between 
atoms is not too small, so that their atomic 
orbitals do not overlap, and, on 
the other hand, the lattice is finite and its 
overall dimensions are smaller 
than the laser wavelength $\lambda$.
However, in most cases, in particular within the NNA,
which is the most common one and used throughout this paper, the sufficient
condition is much less restrictive,
only requiring that the interatomic separation $l_a \ll \lambda$, which is
the same limit as in the standard LLT \cite{BornWolf}.

The LF acting upon a given atom located at a point $\vect{r}$ is a 
superposition of the incident (``external'') field $\vect{E}_\mathrm{in}$
and the sum of the fields $\vect{E}_\mathrm{dp}(\vect{r}, \vect{r}')$ from 
surrounding dipoles at all other lattice positions $\vect{r}'$
over the entire lattice:
\begin{equation}
\vect{E}_\mathrm{L} (\vect{r}) = \vect{E}_\mathrm{in} (\vect{r}) 
+ \sum^{\vect{r}' \ne \vect{r}}_\mathrm{lattice}
    \vect{E}_\mathrm{dp}(\vect{r}, \vect{r}'),
\label{eq-LF-1}
\end{equation}
where the near fields of the surrounding atomic dipoles $\vect{p}(\vect{r}')$ 
are dominated by the non-radiative (quasi-static) components 
\cite{LandauLifshitz},
\begin{equation}
\vect{E}_\mathrm{dp}(\vect{r}, \vect{r}')
= \frac{3\vect{u} [\vect{p}(\vect{r}') \cdot \vect{u}] - \vect{p}(\vect{r}')}%
       {\epsilon |\vect{r}' - \vect{r}|^3}.
\label{eq-LF-2}
\end{equation}
Here $\vect{u} \equiv (\vect{r} - \vect{r}') / |\vect{r} - \vect{r}'|$ 
is a unit vector along the line connecting the two atoms,
$\epsilon$ is the background dielectric constant ($\epsilon = 1$ in vacuum
and $\epsilon \ne 1$ if a host medium is present).
The atomic dipoles $\vect{p}(\vect{r}')$ are, in turn, induced by the LF 
acting upon them.
In the case of linear optical response of a two-level atom, its induced 
dipole moment is
\begin{equation}
\vect{p}(\vect{r}) = - \vect{E}_\mathrm{L} (\vect{r})
\frac{2|d_a|^2}{\hbar\Gamma (\delta + i)},
\label{eq-LF-3}
\end{equation}
where $\delta \equiv T (\omega - \omega_0)$ is a dimensionless laser frequency 
detuning and $T = 2 / \Gamma$ is the transverse relaxation time of the atom 
with a resonant homogeneous linewidth $\Gamma$.
The condition that the electron orbitals of neighboring atoms do not overlap
implies that $l_a \gg | d_a | / e$ and justifies our use of a semiclassical 
approach.
Nonlinear effects in systems of resonant atoms could be included into this 
picture via the saturation nonlinearity of the two-level system.
As we have recently demonstrated \cite{Kaplan2008PRL,Kaplan2009PRA}, the 
nonlinearity enables interesting effects with promising applications, like the
optical bistability and hysteresis.
In the present paper, however, we will only consider \emph{linear} effects
in 2D lattices of resonant atoms.

We can rewrite Eqs.~(\ref{eq-LF-1})--(\ref{eq-LF-3}) in the following closed 
form,
\begin{eqnarray}
\vect{E}_\mathrm{L} (\vect{r}) &=& \vect{E}_\mathrm{in} (\vect{r}) 
- \frac{Q}{4} \sum^{\vect{r}' \ne \vect{r}}_\mathrm{lattice}
\frac{l_a^3 }{ |\vect{r}' - \vect{r}|^3 }
\nonumber
\\[2\jot]
&&\qquad\times
\left\{ 3 \vect{u} [ \vect{E}_\mathrm{L} (\vect{r}') \cdot  \vect{u} ]
   - \vect{E}_\mathrm{L} (\vect{r}') \right\},
\label{eq-LF}
\end{eqnarray}
where we introduced the dimensionless strength $Q$ of the dipole-dipole 
coupling between neighboring atoms, which is easily controlled through the 
normalized frequency detuning $\delta$,
\begin{equation}
Q = \frac{Q_a}{\delta + i}.
\label{eq-Q}
\end{equation}
The maximum normalized strength of interaction between neighboring atoms,
\begin{equation}
Q_a = \frac{8 |d_a|^2}{\epsilon \hbar \Gamma l_a^3},
\label{eq-Q-a}
\end{equation}
is reached at zero detuning of the incident laser frequency from the atomic 
resonance ($\delta = 0$).
Note that this is the point where our approach drastically departs
from the conventional LF theory, 
where the LF is implied quasi-uniform on a scale of many $l_a$, so that 
$\vect{E}_\mathrm{L}(\vect{r}) \approx \vect{E}_\mathrm{L}(\vect{r}')$.
More details on the general LF model that we use in this paper, including some
quantitative estimates of $Q_a$ for realistic systems, can be found in our 
recent publication \cite{Kaplan2009PRA}.

In the present paper we restrict our considerations to the 
\emph{nearest-neighbor approximation} (NNA), in which only interactions between 
the closest atoms are taken into account in Eq.~(\ref{eq-LF}).
As we demonstrated in \cite{Kaplan2008PRL,Kaplan2009PRA}, this approximation 
leads to qualitatively similar results compared to the full solution, that 
takes into account interactions between all pairs of atoms in the system.
The NNA allows us to derive simpler analytic expressions and undertake
numerical simulations for reasonably large lattices.

\section{Locsitons in triangular lattices: in-plane polarization}
\label{sec-in-plane}

In this section we consider the emergence and properties of locsitons in 
infinite unbounded planar lattices of atoms.
Here we aim at studying locsitons 
in their ``simplest'' form, without the 
system boundaries complicating the picture.
At the same time, we will continue to assume the near-field character of the
interatomic interaction, because our ultimate goal is to study systems that 
are much smaller than $\lambda$.
For the same reason, we will assume the incident field to be uniform 
throughout the lattice, 
$\vect{E}_\mathrm{in}(\vect{r}) \equiv \vect{E}_\mathrm{in}$,
although our formalism is not limited to this case.
As we have proven in \cite{Kaplan2008PRL,Kaplan2009PRA} for linear (1D) 
systems of interacting atoms, such an approach is indeed of a great help for 
understanding the locsitons' behavior.

\begin{figure}
\includegraphics[width=3.25in]{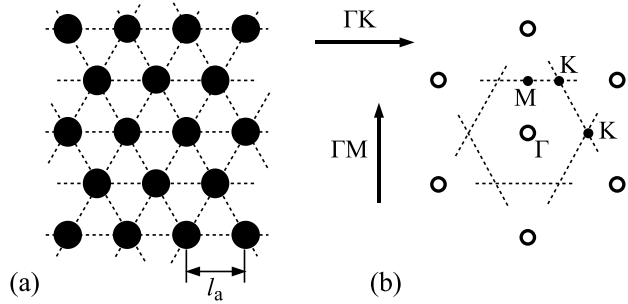}
\caption{%
(a) Geometry of an equilateral triangular lattice of resonant atoms.
(b) The corresponding reciprocal lattice, shown with open circles, and some 
high-symmetry points and directions in the first Brillouin zone.
The dashed lines in (b) illustrate constructing the first Brillouin zone as 
the Wigner-Seitz cell of the reciprocal lattice.}
\label{fig-geom}
\end{figure}

We consider an equilateral triangular lattice (also known as a hexagonal 
lattice) of atoms, which has a six-fold rotation symmetry and belongs to 
the plane symmetry group (wallpaper group) \emph{p6m}.
[See Fig.~\ref{fig-geom}(a).]
This lattice type has a remarkable property of providing the most close-packed 
configuration of identical circular objects in a plane.
Each atom in the lattice  has six neighbors at the distances of $l_a$.
The ``second layer'' of neighbors is removed by $\sqrt{3} l_a$ or $2 l_a$;
considering the fast decrease of the dipole-dipole interaction force with 
the distance ($\propto 1/r^3$), the NNA which ignores interaction with the 
second and further layers of neighbors is expected to work well.

As in the case of 1D arrays of atoms \cite{Kaplan2008PRL,Kaplan2009PRA}, two 
major cases can be studied separately, where the incident laser field 
$\vect{E}_\mathrm{in}$ lies \emph{in plane} (the ``$\parallel$'' case) or is 
\emph{normal} to the plane of the lattice (the ``$\perp$'' case).
The linear optical response in the case of any other incident polarization can 
be obtained using a superposition of these two configurations.
Of the two major cases, the ``$\parallel$'' configuration is by far more 
interesting, 
exhibiting richer locsiton behavior and differing significantly from the 1D 
case due to the crucial in-plane anisotropy of the dipole-dipole interaction.
It is also easier to implement, as the laser beam may be incident at the 
normal to the lattice of atoms, with the incident field being virtually 
uniform on a scale of many wavelengths.
The ``$\perp$'' configuration presents a much closer analogy to the 1D 
problems considered in \cite{Kaplan2008PRL,Kaplan2009PRA}, although some 
details inevitably differ.
We will discuss it below in Sec.~\ref{sec-normal}.
Further in this section, we only consider the ``$\parallel$'' case, where
both $\vect{E}_\mathrm{in}$ and $\vect{E}_\mathrm{L}$ lie in the lattice plain.

\subsection{Near-ring approximation}
\label{subsec-NRA}

Optical response of a triangular lattice of atoms is, in general, anisotropic:
it depends on the orientation of $\vect{E}_\mathrm{in}$ in the lattice plane 
with respect to the symmetry directions of the lattice.
We have demonstrated, however, 
that one can use the near-ring approximation 
(NRA) to describe the behavior of long-wavelength locsitons in the lattice 
\cite{Kaplan2008PRL} in an isotropic fashion.
In this approximation, the contribution from the surrounding dipoles to the LF 
$\vect{E}_\mathrm{L}$ [the second term in Eq.~(\ref{eq-LF-1})] is substituted 
with a field from an effective dipole ring with the radius $l_a$, such that 
the polarizability of the nearest six atoms is evenly redistributed along the 
ring.
The NRA is thus a further simplification of the NNA, making the model 
isotropic in the lattice plane.
Within the NRA we then replace the summation in Eq.~(\ref{eq-LF}) with an 
integration over the imaginary ring, so that the equation for the LF
becomes
\begin{equation}
\vect{E}_\mathrm{L} (\vect{r}) = \vect{E}_\mathrm{in} - \frac{3}{4\pi} Q 
\int_0^{2\pi} \vect{E}_\mathrm{L} (\vect{r} + l_a\vect{u})\, 
(3 \cos^2 \theta - 1)\, d \theta,
\label{eq-LF-NRA}
\end{equation}
where $\vect{u}$ is a unit vector in the direction from the center $\vect{r}$ 
of the effective ring to a point at the ring, and 
$\theta$ is the polar angle of $\vect{u}$ counted from the direction of 
$\vect{E}_\mathrm{in}$ in the lattice plane.
The strength of the dipole-dipole coupling between neighboring atoms is still
given by the dimensionless parameter $Q$ [Eq.~(\ref{eq-Q})].

Equation (\ref{eq-LF-NRA}) has a uniform (\emph{Lorentz}) solution
\begin{equation}
\bar{\vect{E}}_\mathrm{L}
= \frac{\delta + i}{\delta - \delta_\mathrm{LL} + i} \vect{E}_\mathrm{in},
\label{eq-Lorentz}
\end{equation}
where
\begin{equation}
\delta_\mathrm{LL}^\parallel = - \frac{3}{4} Q_a
\label{eq-dLL}
\end{equation}
is the normalized frequency detuning at which the Lorentz-Lorenz resonance
is achieved in the triangular lattice of atoms within the NRA.
As we will see in Sec.~\ref{sec-normal}, $\delta_\mathrm{LL}$ is different 
for the ``$\parallel$'' and ``$\perp$'' configurations, but to simplify the 
formulas, we omit the index~``$\parallel$'' 
everywhere in this section and in Sec.~\ref{sec-finite}, except for 
Eq.~(\ref{eq-dLL}).
From Eq.~(\ref{eq-Lorentz}) one can see that, if 
$|\delta_\mathrm{LL}| \sim Q_a \gg 1$, the uniform Lorentz LF is suppressed
when the laser is tuned closely to the exact atomic resonance, 
$\delta \approx 0$, reaching its minimum intensity
\begin{equation}
|\bar{\vect{E}}_\mathrm{L}|_\mathrm{min}^2
\approx \frac{|\vect{E}_\mathrm{in}|^2}{1 + \delta_\mathrm{LL}^2}.
\label{eq-Ebar-min}
\end{equation}
The LF in this case is effectively ``pushed out'' by the lattice atoms.
Interestingly, a huge LF enhancement is reached at a red-shifted frequency,
at $\delta \approx \delta_\mathrm{LL} < 0$, where
\begin{equation}
|\bar{\vect{E}}_\mathrm{L}|_\mathrm{max}^2
\approx (1 + \delta_\mathrm{LL}^2)\, |\vect{E}_\mathrm{in}|^2.
\label{eq-Ebar-max}
\end{equation}

Note that Eqs.~(\ref{eq-Lorentz}), (\ref{eq-Ebar-min}), 
and~(\ref{eq-Ebar-max}) are very similar to the corresponding equations for 
1D arrays of atoms which were discussed in \cite{Kaplan2008PRL,Kaplan2009PRA}.
The only difference is that the relation between $\delta_\mathrm{LL}$ and 
$Q_a$ there was 
$\delta_\mathrm{LL} = - Q_a$ if $\vect{E}_\mathrm{in}$ is parallel to the 
array (and dipoles are aligned ``head-to-tail'')
and $\delta_\mathrm{LL} = Q_a/2$ if $\vect{E}_\mathrm{in}$ is perpendicular to 
the array (and dipoles are aligned ``side-to-side''), assuming the NNA
[see, e.\,g., Eq.~(3.4) of Ref.~\cite{Kaplan2009PRA}].
Here, within the NRA, $\delta_\mathrm{LL}$ does not depend on the incident 
field polarization in the lattice plane.
Quite naturally, its value (\ref{eq-dLL}) lies in-between the two values for 
the 1D array, because mutual orientations of different pairs of dipoles 
in the lattice vary between the two extremes.

\begin{figure}
\includegraphics[width=2.9in]{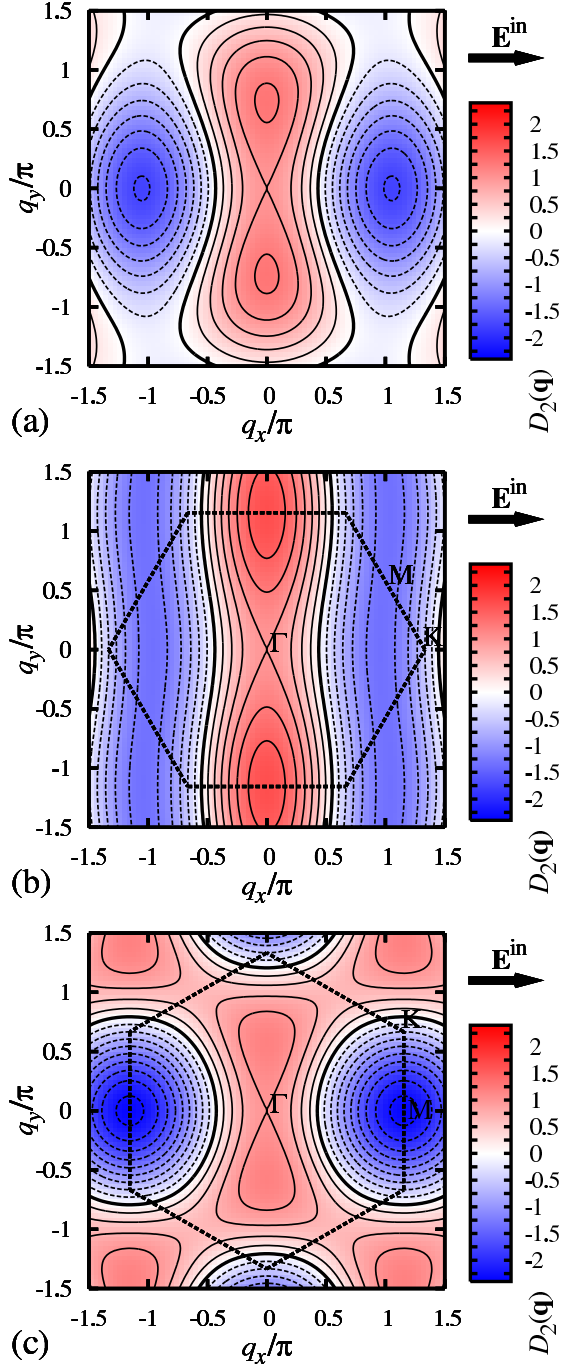}
\caption{%
(Color online) Dispersion dependences for locsitons in a triangular lattice of 
atoms (a) in the near-ring approximation;
(b) for $\vect{E}_\mathrm{in} \parallel \Gamma\mathrm{K}$;
(c) for $\vect{E}_\mathrm{in} \parallel \Gamma\mathrm{M}$.
Isolines of equal $D_2(\vect{q})$ are spaced at 0.25 (i.\,e., they correspond
to the ticks on the color bars);
solid isolines and red shading correspond to positive $D_2(\vect{q})$, 
dashed isolines and blue shading correspond to negative ones,
while the thicker solid isoline marks the zero level.
The hexagons in plots (b,c) show the boundaries of the first Brillouin zone.}
\label{fig-disp-2D}
\end{figure}

The frequency dependence of a \emph{spatially uniform} LF, like in 
Eq.~(\ref{eq-Lorentz}), and the associated Lorentz shift are long known 
phenomena.
Similar effect was also observed experimentally in \emph{alkali} vapors 
\cite{Maki1991}.
The unusual new phenomenon is that
in ordered low-dimensional structures there are 
\emph{spatially varying} solutions, which we call \emph{locsitons} 
\cite{Kaplan2008PRL}, 
that emerge at some values of $Q$ in addition to the uniform LF.
We will look for the locsitons in the form 
of 2D plane-wave excitations of the LF:
\begin{equation}
\Delta\vect{E}_\mathrm{L} 
\propto \exp ( i \vect{q} \cdot \vect{r} / l_a ),
\label{eq-DeltaE}
\end{equation}
where $\vect{q}$ is the normalized wave vector of the locsiton.
By substituting 
$\vect{E}_\mathrm{L} = \bar{\vect{E}}_\mathrm{L} + \Delta\vect{E}_\mathrm{L}$
into Eq.~(\ref{eq-LF-NRA}) we obtain the dispersion relation
for the wave vector $\vect{q}$ in an integral form:
\begin{equation}
1 + \frac{3Q}{4\pi} \int_0^\pi (3 \cos 2\theta + 1) 
  \cos[q\cos(\theta - \psi)]\, d\theta = 0,
\label{eq-disp-NRA-1}
\end{equation}
where $\psi$ is the polar angle of $\vect{q}$ counted from the direction of 
$\vect{E}_\mathrm{in}$ in the lattice plane.
Using the standard expansion of trigonometric functions
with a harmonic argument into Bessel functions
[see, e.\,g., Eq.~(21.8-25\textit{a}) in Ref.~\cite{KornKorn}],
\begin{equation}
\cos(q\sin\phi) = J_0(q) + 2 \sum_{m=1}^\infty J_{2m}(q) \cos(2m\phi),
\label{eq-Bessel-exp}
\end{equation}
where $J_m(q)$ is the Bessel function of the first kind,
we evaluate the integral in Eq.~(\ref{eq-disp-NRA-1}) 
in an explicit form and write Eq.~(\ref{eq-disp-NRA-1}) as
\begin{equation}
1 + \frac{3}{4} Q\, [J_0(q) - 3 J_2(q) \cos(2\psi)] = 0.
\label{eq-disp-NRA-2}
\end{equation}
By substituting $Q$ from Eqs.~(\ref{eq-Q}) and (\ref{eq-dLL}), we can rewrite 
the last formula as an explicit dispersion relation 
connecting the normalized detuning 
$\delta$ with the normalized locsiton wave vector $\vect{q}$ which is 
represented by its polar coordinates $q$ and $\psi$:
\begin{equation}
D_2^\mathrm{NRA}(\vect{q}) \equiv J_0(q) - 3 J_2(q) \cos(2\psi)
= \frac{\delta + i}{\delta_\mathrm{LL}}.
\label{eq-disp-NRA-3}
\end{equation}
In the limit of low absorption, 
i.\,e., $| \delta_\mathrm{LL} | \gg 1$, the r.\,h.\,s.\ of 
Eq.~(\ref{eq-disp-NRA-3}) may be replaced with $\delta / \delta_\mathrm{LL}$.
A more detailed analysis of the dispersion relation shows that spatial 
oscillations---locsitons with almost real~$\vect{q}$---emerge in a 
limited frequency band around $\omega_0$ with a normalized bandwidth 
$\sim |\delta_\mathrm{LL}|$.
It is exactly the frequency range where the interaction between resonant 
atomic dipoles may become comparable to or much stronger than the effects of 
the external field $\vect{E}_\mathrm{in}$.
The dispersion dependence 
$\delta / \delta_\mathrm{LL} = D_2^\mathrm{NRA}(\vect{q})$ in this case is 
shown in Fig.~\ref{fig-disp-2D}(a), where the external field is assumed to be 
aligned with the $x$ axis.
At every given laser frequency $\omega$, locsitons with a whole range of wave 
vectors $\vect{q}$ may be excited.
This set of~$\vect{q}$, all having different orientations, is represented by 
an isoline for the respective
$\delta / \delta_\mathrm{LL} = D_2^\mathrm{NRA}(\vect{q})$
in Fig.~\ref{fig-disp-2D}(a).
Recalling that $\delta_\mathrm{LL} < 0$ [see Eq.~(\ref{eq-dLL})], we may 
notice that solid isolines (red shaded areas) correspond to red-shifted laser 
frequencies ($\omega < \omega_0$), while dashed isolines (blue shaded areas) 
correspond to blue-shifted frequencies.
The non-circular shape of the isolines reflects the highly anisotropic 
dispersion dependence for locsitons in the 2D lattice
in the case of short-wave locsitons.
In particular, at any given normalized frequency detuning $\delta$, locsitons 
with different orientations of $\vect{q}$ with respect to 
$\vect{E}_\mathrm{in}$ may have very different $q$ and, consequently, 
different wavelengths.
At the same time, because the NRA describes the lattice in an averaged way, 
the orientation of $\vect{E}_\mathrm{in}$ with respect to the lattice does not 
affect the result.

In the general case, especially when looking at the limitations
on the size of the structure,
one needs to consider complex locsiton wave vectors 
$\vect{q} = \vect{q}' + i\vect{q}''$ in the dispersion relations
(\ref{eq-disp-NRA-1}) or (\ref{eq-disp-NRA-3}),
like it was done for 1D arrays of resonant atoms in \cite{Kaplan2009PRA}.
This would allow to describe the dissipation of 2D locsitons, which is most 
prominent near the Lorentz resonance, and \emph{evanescent} locsitons, 
which exist outside the locsiton band. 
One can also calculate the group velocity of the locsiton, 
$\vect{v}_\mathrm{gr} = (l_a/T)(d\delta/d\vect{q})$, by taking a derivative of 
Eq.~(\ref{eq-disp-NRA-3}).
Due to the anisotropy of Eq.~(\ref{eq-disp-NRA-3}), one may expect that the 
resulting dissipation and group velocity are highly dependent on the 
orientation of $\vect{q}$ with respect to $\vect{E}_\mathrm{in}$.

By its nature, the NRA only works well for long-wave locsitons, i.\,e., those 
with relatively small $q$.
Indeed, for a long-wave locsiton, the LF at neighboring atoms differs 
insignificantly, so one may reasonably expect that replacing the neighboring 
dipoles at their actual positions with an effective dipole ring will not cause
a significant change to the result.
Our results obtained without resorting to the NRA (see Sec.~\ref{subsec-BZ}), 
suggest that the NRA provides a good quantitative description of locsitons for 
$q \lesssim \pi/2$, while it gives reasonable \emph{qualitative} estimates for 
$q$ up to $\sim \pi$.
One may use the two first terms from the Taylor series expansion to analyze 
Eq.~(\ref{eq-disp-NRA-3}) at small $q$, which is easy to do by recalling that 
$J_0(q) = 1 - \frac{1}{4} q^2 + O(q^4)$ and
$J_2(q) = \frac{1}{8} q^2 + O(q^4)$
[cf.\ Eq.~(21.8-3) in Ref.~\cite{KornKorn}], so that
\begin{equation}
D_2^\mathrm{NRA}(\vect{q}) 
\approx 1 - \frac{1}{8} q^2 [2 + 3 \cos(2\psi)] + O(q^4).
\label{eq-disp-NRA-small-q}
\end{equation}
In the long-wavelength limit, $D_2^\mathrm{NRA}(\vect{q} = \vect{0}) = 1$ is 
reached at $\delta = \delta_\mathrm{LL}$, yielding the uniform, Lorentz, 
solution for the LF.
The second term in the r.\,h.\,s.\ of Eq.~(\ref{eq-disp-NRA-small-q}) 
indicates that the anisotropy with respect to the locsiton polarization shows 
up even at very small $q$, where Eqs.~(\ref{eq-disp-NRA-3}) 
and~(\ref{eq-disp-NRA-small-q}) can be rewritten as
\begin{equation}
q^2 = -\frac{8}{2 + 3 \cos(2\psi)} 
    \frac{(\delta - \delta_\mathrm{LL}) + i}{\delta_\mathrm{LL}}.
\label{eq-disp-NRA-small-q-2}
\end{equation}

Note that in describing locsitons at very small $q$ (i.\,e., for 
$\delta \approx \delta_\mathrm{LL}$), the imaginary unit in the last formula 
cannot always be neglected compared to $\delta - \delta_\mathrm{LL}$.
The locsiton dissipation in this case may become significant, depending on 
$\delta_\mathrm{LL}$, and the necessity of considering complex~$\vect{q}$ 
could appreciably complicate calculations.
We will not go into the details of this special case in the present paper.

In 1D arrays of atoms, the Lorentz resonance, 
$\delta = \delta_\mathrm{LL}$, always coincides with
one of the edges of the locsiton 
frequency band \cite{Kaplan2008PRL,Kaplan2009PRA}.
In that case, when one tunes the frequency 
of the incident laser beam towards 
$\omega_0$ from the side of the Lorentz resonance, locsitons with 
the longest wavelength emerge first very close to the Lorentz resonance. 
The locsiton wavelength subsequently decreases as we approach and move past 
$\omega_0$.
In a 2D lattice, the Lorentz resonance lies \emph{within} the locsiton band,
which means that, when similarly tuning the laser frequency towards 
$\omega_0$, short-wave locsitons will be excited first.
In particular, locsitons with $\psi \approx \pm\pi/2$ (i.\,e., nearly 
transverse locsitons) may be exited at $\delta / \delta_\mathrm{LL} > 1$, and 
can be thus viewed as the ``easiest to excite'' on the Lorentz side of the 
band.
On the opposite, ``anti-Lorentz'' side of the band, where 
$\delta / \delta_\mathrm{LL} \lesssim -1$, nearly longitudinal locsitons
with $\psi \approx 0$~or~$\pi$ lie closer to the band edge and thus are easier 
to excite.
The exact positions of the edges of the locsiton band cannot be found within 
the NRA, because minima and maxima of $D_2^\mathrm{NRA}(\vect{q})$ are reached 
at such $\vect{q}$ where the NRA may only be used for qualitative estimates.
The case of larger $q$ is addressed when we go beyond the NRA in the next 
subsection.

\subsection{Locsitons in the first Brillouin zone}
\label{subsec-BZ}

When going beyond the NRA, the orientation of $\vect{E}_\mathrm{in}$ within 
the lattice plane becomes an important factor, except for small $q$.
Staying within the NNA, i.\,e., only taking into account the six nearest 
neighbors in Eq.~(\ref{eq-LF}) (but \emph{individually}, instead
of them being washed out over the ring, as in the NRA), 
we are still able to approach the problem 
analytically.
The resulting equation is
\begin{eqnarray}
\vect{E}_\mathrm{L} (\vect{r}) &=& \vect{E}_\mathrm{in}
- \frac{Q}{4} \sum_{\vect{u}_\mathrm{K}}
\left\{ 3 \vect{u}_\mathrm{K} 
        [ \vect{E}_\mathrm{L} (\vect{r} + l_a\vect{u}_\mathrm{K}) 
          \cdot \vect{u}_\mathrm{K} ]
\right.\nonumber
\\[2\jot]
&&\left.\qquad\qquad\qquad{}
   - \vect{E}_\mathrm{L} (\vect{r} + l_a\vect{u}_\mathrm{K}) \right\},
\label{eq-LF-NNA}
\end{eqnarray}
where $\vect{u}_\mathrm{K}$ denotes any of the six unit vectors pointing in 
the directions from the atom (located at $\vect{r}$) to one of its nearest 
neighbors.

The uniform, Lorentz, solution of Eq.~(\ref{eq-LF-NNA}) is still given by 
Eq.~(\ref{eq-Lorentz}) and~(\ref{eq-dLL}), which supports the above-mentioned
convergence of the NRA and NNA results at $q \to 0$.
Spatially varying \emph{locsiton} solutions are found as in the previous 
subsection by using the ansatz (\ref{eq-DeltaE}) in Eq.~(\ref{eq-LF-NNA}).
The corresponding dispersion relation $\delta(\vect{q})$ for locsitons in a 2D 
triangular lattice can be now written as
\begin{equation}
D_2^\mathrm{NNA}(\vect{q}) 
\equiv \sum_{n=0}^2 \left( \cos 2\theta_n + \frac{1}{3} \right) 
    \cos[q \cos(\theta_n - \psi)]
= \frac{\delta + i}{\delta_\mathrm{LL}}.
\label{eq-disp-NNA}
\end{equation}
where $\theta_n = \theta_0 + n\pi/3$, and $\vect{q}$ is represented by its 
polar coordinates $q$ and $\psi$.
The orientation of the lattice with respect to the incident field 
$\vect{E}_\mathrm{in}$ is described by $\theta_0$, which is the angle that 
one of the vectors $\vect{u}_\mathrm{K}$ makes with $\vect{E}_\mathrm{in}$
(the fact that $\theta_0$ is not unique does not affect the result).
The ultimate proof that the results of the NRA and the more precise NNA 
converge for long locsiton wavelengths can be obtained by taking a Taylor 
series expansion of Eq.~(\ref{eq-disp-NNA}) at $q \to 0$.
By only retaining terms up to the order of $q^2$ and calculating all the 
necessary sums and products of trigonometric functions, we find that
Eqs.~(\ref{eq-disp-NRA-small-q}) and~(\ref{eq-disp-NRA-small-q-2})
still hold in the NNA.

It is sufficient to find locsitons with $\vect{q}$ lying within the first 
Brillouin zone of the reciprocal lattice, because any solutions with 
$\vect{q}$ lying outside the first Brillouin zone are physically equivalent to 
them due to the discrete nature of our system.
The dispersion relation $D_2^\mathrm{NNA}(\vect{q})$ in 
Eq.~(\ref{eq-disp-NNA}) also has the required symmetry and periodicity.
It is common to use high-symmetry points in the first Brillouin zone to denote 
most interesting directions in the lattice [see Fig.~\ref{fig-geom}(b)].
Note that in terms of $\vect{u}_\mathrm{K}$ we may write 
$\Gamma\mathrm{K} \parallel \vect{u}_\mathrm{K}$ and
$\Gamma\mathrm{M} \perp \vect{u}_\mathrm{K}$.

The dispersion dependence 
$\delta / \delta_\mathrm{LL} = D_2^\mathrm{NNA}(\vect{q})$ in the case of 
$| \delta_\mathrm{LL} | \gg 1$ is shown in Fig.~\ref{fig-disp-2D}(b,c)
for two different orientations of $\vect{E}_\mathrm{in}$ with respect to the 
lattice:  in Fig.~\ref{fig-disp-2D}(b) $\theta_0 = 0$ 
(i.\,e., $\vect{E}_\mathrm{in} \parallel \Gamma\mathrm{K}$), 
while in Fig.~\ref{fig-disp-2D}(c) $\theta_0 = \pi/2$
(i.\,e., $\vect{E}_\mathrm{in} \parallel \Gamma\mathrm{M}$).
The $x$ axis on the plots is aligned with the external field 
$\vect{E}_\mathrm{in}$;
the hexagonal boundaries of the first Brillouin zones are shown with thicker 
dashed lines.
The central parts of all three plots in Fig.~\ref{fig-disp-2D} are very 
similar, which reflects our finding that Eqs.~(\ref{eq-disp-NRA-small-q}) 
and~(\ref{eq-disp-NRA-small-q-2}) hold for long locsiton wavelengths within 
both NRA and NNA.
At the same time, significant differences accumulate closer to the boundaries 
of the first Brillouin zone.
Fig.~\ref{fig-disp-2D}(b,c) also show that for both orientations of 
$\vect{E}_\mathrm{in}$ the maxima and minima of $D_2^\mathrm{NNA}(\vect{q})$ 
are attained at the zone boundaries, specifically, at different M points.

\begin{figure}
\includegraphics[width=3.25in]{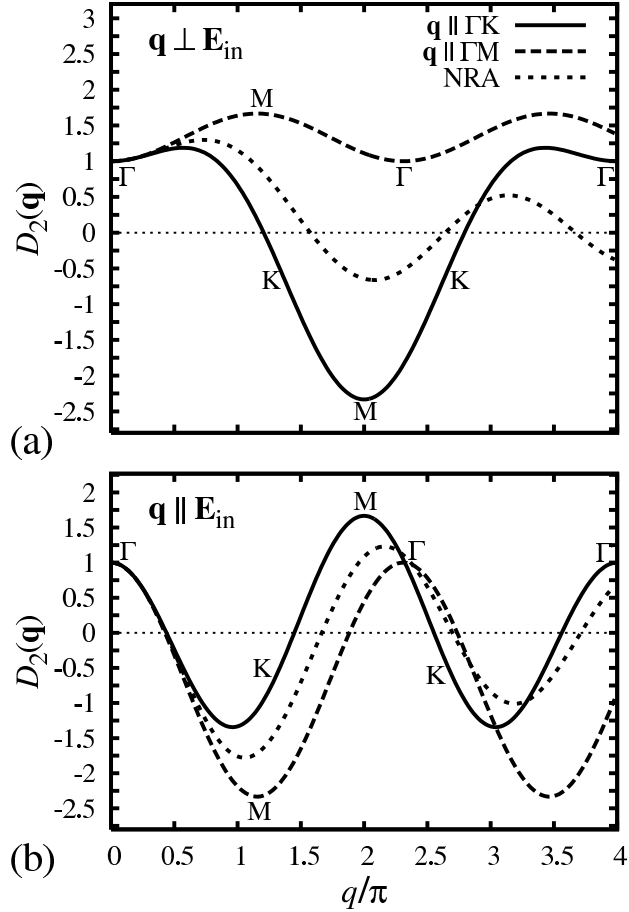}
\caption{%
Dispersion dependences for
(a) transverse locsitons ($\vect{q} \perp \vect{E}_\mathrm{in}$) and 
(b) longitudinal locsitons ($\vect{q} \parallel \vect{E}_\mathrm{in}$)
in a triangular lattice of atoms.
Different curves correspond to the near-ring approximations (NRA) and
the NNA for two different orientations of $\vect{q}$ with respect to the 
lattice.}
\label{fig-disp-1D}
\end{figure}

To facilitate the comparison of the three plots in Fig.~\ref{fig-disp-2D} and 
finding the edges of the locsiton band, the NNA dispersion dependencies 
for two high-symmetry directions are presented in Fig.~\ref{fig-disp-1D}
and compared to the the NRA result.
For both longitudinal and transverse locsitons, Fig.~\ref{fig-disp-1D}
shows corresponding cross-sections of the three plots of 
Fig.~\ref{fig-disp-2D}.
As we already noted earlier in this paper, the NRA result is only meaningful 
for $q$ up to $\sim\pi$; 
Fig.~\ref{fig-disp-1D} suggests that it is a good approximation for the NNA 
result at $q \lesssim \pi/2$, regardless of the orientation of 
$\vect{E}_\mathrm{in}$ with respect to the lattice.
It is instructive to give explicit analytic expressions for each of the 
four NNA-based dependencies in Fig.~\ref{fig-disp-1D}.
The respective dispersion relations derived from Eq.~(\ref{eq-disp-NNA}) and 
the corresponding ranges for $\delta / \delta_\mathrm{LL}$, obtained in the 
assumption that $|\delta_\mathrm{LL}| \gg 1$, are as follows:

\begin{itemize}
\item[(a)]
$\vect{E}_\mathrm{in} \parallel \Gamma\mathrm{K}$,
$\vect{q} \parallel \Gamma\mathrm{M} \perp \vect{E}_\mathrm{in}$ 
[long-dashed line in Fig.~\ref{fig-disp-1D}(a)],
\begin{equation}
\cos\frac{q\sqrt{3}}{2} = 4 - 3\frac{\delta + i}{\delta_\mathrm{LL}},
\label{eq-disp-2D-a}
\end{equation}
\begin{equation}
1 \le \frac{\delta}{\delta_\mathrm{LL}} \le 1\frac{2}{3};
\label{eq-range-a}
\end{equation}
\item[(b)]
$\vect{E}_\mathrm{in} \parallel \Gamma\mathrm{M}$,
$\vect{q} \parallel \Gamma\mathrm{K} \perp \vect{E}_\mathrm{in}$
[solid line in Fig.~\ref{fig-disp-1D}(a)],
\begin{equation}
\cos\frac{q}{2} = \frac{1}{8} 
    \left(5 \pm \sqrt{57 - 48\frac{\delta + i}{\delta_\mathrm{LL}}}\right),
\label{eq-disp-2D-b}
\end{equation}
\begin{equation}
-2\frac{1}{3} \le \frac{\delta}{\delta_\mathrm{LL}} \le 1\frac{3}{16},
\label{eq-range-b}
\end{equation}
where the minimum of $\delta / \delta_\mathrm{LL}$ is reached outside the 
first Brillouin zone; 
this point more appropriately belongs to the case (d) below;
\item[(c)]
$\vect{E}_\mathrm{in} \parallel \Gamma\mathrm{K}$,
$\vect{q} \parallel \Gamma\mathrm{K} \parallel \vect{E}_\mathrm{in}$
[solid line in Fig.~\ref{fig-disp-1D}(b)],
\begin{equation}
\cos\frac{q}{2} = \frac{1}{16} \left[1 \pm 
\sqrt{1 + 32 \left(4 + 3\frac{\delta + i}{\delta_\mathrm{LL}}\right)}\right],
\label{eq-disp-2D-c}
\end{equation}
\begin{equation}
-1\frac{33}{96} \le \frac{\delta}{\delta_\mathrm{LL}} \le 1\frac{2}{3};
\label{eq-range-c}
\end{equation}
where the maximum of $\delta / \delta_\mathrm{LL}$ is reached outside the 
first Brillouin zone, which point more appropriately belongs to the case (a);
\item[(d)]
$\vect{E}_\mathrm{in} \parallel \Gamma\mathrm{M}$,
$\vect{q} \parallel \Gamma\mathrm{M} \parallel \vect{E}_\mathrm{in}$
[long-dashed line in Fig.~\ref{fig-disp-1D}(b)],
\begin{equation}
\cos\frac{q\sqrt{3}}{2} 
    = \frac{1}{5} \left(2 + 3 \frac{\delta + i}{\delta_\mathrm{LL}}\right),
\label{eq-disp-2D-d}
\end{equation}
\begin{equation}
-2\frac{1}{3} \le \frac{\delta}{\delta_\mathrm{LL}} \le 1.
\label{eq-range-d}
\end{equation}
\end{itemize}

The inequalities (\ref{eq-range-b}) and~(\ref{eq-range-c}) also represent the 
edges of the locsiton bands for the cases of 
$\vect{E}_\mathrm{in} \parallel \Gamma\mathrm{M}$ and
$\vect{E}_\mathrm{in} \parallel \Gamma\mathrm{K}$, respectively
(which explains our desire not to restrict the range of $q$ to the first 
Brillouin zone when obtaining these inequalities).
Note that locsitons with the largest red shift 
($\delta/\delta_\mathrm{LL} > 0$) can be achieved with 
$\vect{E}_\mathrm{in} \parallel \Gamma\mathrm{K}$, while 
locsitons with the largest blue shift ($\delta/\delta_\mathrm{LL} < 0$) can be 
achieved with $\vect{E}_\mathrm{in} \parallel \Gamma\mathrm{M}$.
Therefore, re-orienting the lattice with respect to the polarization of the 
incident laser beam may assist in controlling the type of locsitons excited 
in the lattice.

\section{Finite lattices: in-plane polarization}
\label{sec-finite}

In this section, like in Sec.~\ref{sec-in-plane}, we only consider the case 
where $\vect{E}_\mathrm{in}$ is spatially uniform and lies in the lattice 
plane (the ``$\parallel$'' case), which is easier to achieve if the laser beam 
is incident normally to the lattice plane.
The presence of boundaries and defects in 2D lattices of resonant atoms can 
cause various locsitonic effects, including giant LF resonances, 
formation of dipole strata, and ``magic'' cancellation of the resonant 
LF suppression.
These effects are similar in their nature to their counterparts in 1D arrays 
of resonant atoms \cite{Kaplan2008PRL,Kaplan2009PRA}, but their manifestations 
are much more diverse because of the inherent anisotropy of the dipole-dipole 
interaction in the 2D case, especially in the ``$\parallel$'' geometry.
Because of this, an all-encompassing study of finite 2D lattices is hardly 
possible within this pilot study on the subject,
so we will restrict ourselves to providing some of the most characteristic 
results, which emphasize distinctions from the 1D problem.

\subsection{Size-related resonances and local-field patterns}
\label{subsec-patterns}

Locsitons in a finite 1D array of atoms exhibit size-related resonances, 
characterized by large increases in their amplitudes
at certain frequencies within the locsitonic band,
because locsitons are reflected at the boundaries and form 
standing waves (\emph{strata})
\cite{Kaplan2008PRL,Kaplan2009PRA}.
Essentially, these resonances correspond to locsiton eigenmodes 
defined by the boundaries.
For the \emph{long-wave strata} they are 
similar to oscillations of a quantum particle in a box, 
as, e.\,g., for 1D-confined electrons \cite{Sandomirskii1967,Chernyak2001},
or a common violin string.
It is natural to expect such resonances and eigenmodes
to also exist in higher dimensions, in particular, in 
finite 2D lattices, where
we also encounter locsiton reflections at the boundaries.
An important distinction of the 2D case is that the wave vector $\vect{q}$ of 
a locsiton may have an arbitrary orientation in the lattice plane with 
respect to the incident field $\vect{E}_\mathrm{in}$.
Multiple reflections and interference of locsitons with all possible 
$\vect{q}$ quickly make the whole picture very complicated and highly 
susceptible to minor changes to the size and shape of the lattice patch.
We found that at certain geometries only a limited number of locsiton
eigenmodes are dominant. 
Their interference produces various dipole patterns and strata;
some of them are reminiscent to ``quantum carpets'' \cite{Kaplan2000}.
An important issue is, therefore, how one can control the locsiton patterns 
via the geometry of the lattice patch and the frequency and polarization of 
the laser beam.

One way to engineer a distinct 2D locsiton pattern is to start with a 
rectangular lattice patch and ensure that size-related resonances are achieved 
for locsitons with wave vectors parallel to its boundaries.
We have to choose the lattice shape, such that the size-related
locsiton resonances emerge 
in both dimensions at the same frequency detuning $\delta$.
To simplify our task, we will consider long-wavelength locsitons, which are
not too sensitive to the system sizes and thus are easier to control, and, 
incidentally, also form more pronounced patterns and are described by the 
simpler formula (\ref{eq-disp-NRA-small-q-2}).

In the limit of long-wavelength locsitons ($q \ll 1$, 
$\delta \approx \delta_\mathrm{LL}$), the dispersion 
relations in the cases (a) and (b) described in Sec.~\ref{subsec-BZ}
coincide with each other:
\begin{equation}
q^2 = 8 \frac{(\delta - \delta_\mathrm{LL}) + i}{\delta_\mathrm{LL}}
\label{eq-small-q-ab}
\end{equation}
[cf.\ Eq.~(\ref{eq-disp-NRA-small-q-2}) at $\psi = \pi/2$].
In a similar manner, one obtains approximate solutions for the cases 
(c) and (d), for which $\psi = 0$:
\begin{equation}
q^2 = -\frac{8}{5}
    \frac{(\delta - \delta_\mathrm{LL}) + i}{\delta_\mathrm{LL}}.
\label{eq-small-q-cd}
\end{equation}
By combining the cases (a) and (b) or the cases (c) and (d), we can achieve 
\emph{simultaneous} size-related resonances, i.\,e., excitations
of eigenmodes in both orthogonal directions in a patch 
of the 2D triangular lattice \emph{at the same laser frequency},
if the patch is approximately square in shape.
Resonances of the same order are hereby attained for locsitons with wave 
vectors pointing in the two orthogonal directions;
a sufficient ``squareness'' of the lattice patch can be achieved by choosing 
its size (i.\,e., the numbers of atoms in the two directions).
Locsitons with shorter wavelengths or with wave vectors pointing in different 
directions will be also present, but they will have no significant influence 
on the emerging dipole pattern due to their nonresonant nature.

\begin{figure}
\includegraphics[width=3.25in]{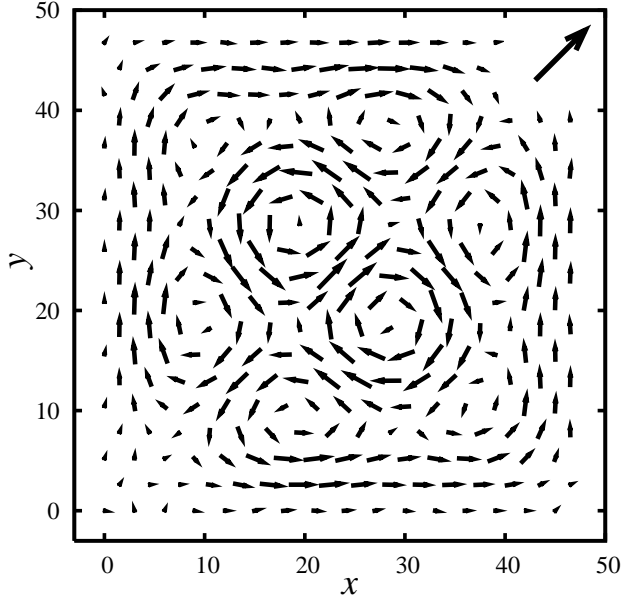}
\caption{%
Vortices in the distribution of the local field in a nearly square patch of a 
2D triangular lattice of atoms at $\delta = -1000$ and 
$\delta_\mathrm{LL} = -987.375$.
To avoid overcrowding of the plot, only one of each nine dipoles is shown.
The incident light wave is polarized in the lattice plane along the diagonal 
of the lattice patch, its field is shown with a big arrow.}
\label{fig-vortices}
\end{figure}

Fig.~\ref{fig-vortices} shows vector patterns that are formed by the 
atomic dipoles induced by the LF.
The atoms are arranged in a $48 \times 56$ patch of an equilateral 
triangular lattice, which results in approximately equal sides of the patch.
The field of the incident electromagnetic wave is uniform and polarized along 
the diagonal of the patch
and its frequency is close to the electronic resonance of the two-level atom.
The frequency of the incident wave is so chosen that the third size-related
resonance (in the order of increasing wavenumbers, counting only those resonances 
allowed by the symmetry of the problem; for more detail
see  \cite{Kaplan2009PRA}) is excited in each dimension.
Eight distinct vortices of the LF are visible in the plot.
Fig.~\ref{fig-vortices} only shows the imaginary parts of the complex field 
amplitudes, because they are dominant for each of the resonant locsitons.
(We would like to note that a pair of vortices,
apparently consistent with 2D-locsiton patterns
originated by the 1-st locsiton resonance in our classification,
was very recently observed in numerical simulations
of plasmonic excitations in a 2D lattice of small
metallic particles \cite{Karpov}.)

\subsection{``Magic shapes''}
\label{subsec-magic}

As we noted in Sec.~\ref{subsec-NRA}, the LF is ``pushed out'' of the 
lattice of strongly interacting dipoles at the exact atomic resonance 
[see Eq.~(\ref{eq-Ebar-min})].
This effect, which we call the resonant LF suppression, represents a 
typical LF behavior at the atomic resonance;
it is not limited to 2D lattices, but also occurs in 1D arrays of interacting 
atoms \cite{Kaplan2008PRL,Kaplan2009PRA} and in many finite 2D structures.
We have shown earlier that in the 1D case, if a linear array of atoms is of a 
certain ``magic size'', one encounters a \emph{cancellation} of the resonant 
LF suppression, where one of the size-related locsitonic resonances 
partially restores the LF in the system \cite{Kaplan2008PRL,Kaplan2009PRA}.

\begin{figure}
\includegraphics[width=3.25in]{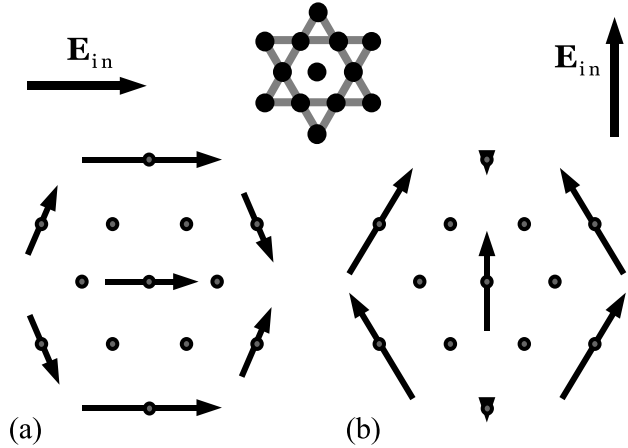}
\caption{%
``Magic'' planar system of 13 resonant atoms for two different orientations 
of $\vect{E}_\mathrm{in}$ shown in plots (a) and (b).
The inset illustrates the geometry of the system.
\label{fig-star}
}
\end{figure}

Finite 2D lattices and similar small systems of resonant atoms provide 
especially interesting examples of cancellation of the resonant LF 
suppression.
Unlike in 1D arrays of atoms, the ``restoration'' of the LF in such 
systems at $\delta = 0$, compared to that in the uniform, Lorentz, case,
can be more complete (up to 100\%).
Like in the 1D case \cite{Kaplan2008PRL,Kaplan2009PRA}, the 2D 
``magic shapes'' have a certain ``cabbalistic'' streak.
For example, in the NNA the effect is most pronounced only in a system of 
$N = 13$ atoms arranged as an equilateral six-point star with an atom at the 
center, for which the maximum restoration of the LF is reached, 
$E_\mathrm{max} / E_\mathrm{in} \approx 1.02$.
The directions and relative amplitudes of the LF at the atoms in this 
system are shown in Fig.~\ref{fig-star}(a) for 
$\vect{E}_\mathrm{in} \parallel \vect{u}_\mathrm{K}$ and
in Fig.~\ref{fig-star}(b) $\vect{E}_\mathrm{in} \perp \vect{u}_\mathrm{K}$.
It is very notable that the system is ``magic'' for both orientations of the 
incident field.
One can see from the picture that the LF is concentrated on the outermost 
atoms and the one at the center, while the LF at the inner hexagon of atoms is 
almost completely suppressed.
This suppression is a manifestation of a
special case of a locsiton standing eigenwave 
in a finite discrete atomic structure.
(Within the NNA, the LF zeroes out at the ``empty'' atomic
locations outside the outer hexagon,
so locsitons make a 2D standing wave.)
In general, the zeroes (nodes) of that wave
are located somewhere in between atoms,
so at each individual atom we have a nonzero LF amplitude.
However, in the magic atomic configuration 
at the precise atomic resonance, $\delta = 0$,
these nodes nearly coincide
with the locations of the inner-hexagon atoms.
Thus, we have a nearly ideal picture of
a 2D standing wave, with large LF intensities
at the antinodes (maxima), located at the central atom and the outer hexagon
atoms, on one hand, and nodes (zeroes), located
at the inner-hexagon atoms, on the other hand.
To an extent, this situation is reminiscent of a
2D standing wave on a water surface in a round bucket with the
first antinode at the center of the bucket, 
where some middle observation points are located 
at the nodes of the wave.
Any symmetry distortion in this system (e.\,g., by attaching a foreign atom or 
molecule to it) would break the balance of the local fields in the system and 
bring back the resonant LF suppression, which is cancelled in the symmetric 
``magic system''.
This effect could potentially lead to
designing nanometer-scale sensors for detecting 
various biological molecules, etc. 
For example, such a nanodevice may include target-specific receptor molecules 
that form a locsiton-supporting ``magic'' system. 
A localized locsiton then would get suppressed whenever a target biomolecule 
attaches to a receptor, otherwise the locsiton suppression at the electronic 
resonance would be ``magically'' cancelled.

\section{Locsitons in triangular lattices: normal polarization}
\label{sec-normal}

The ``$\perp$'' configuration, where the incident field is polarized at the 
normal to the lattice plane, may be realized, for example, by creating a 
standing wave by two counter-propagating laser beams with the beam axes lying 
in the lattice plane.
For systems much smaller than the laser wavelength, the incident field may be 
then assumed nearly uniform. 
Locsitons emerging in the ``$\perp$'' configuration are similar in many 
respects to 1D locsitons discussed in detail in \cite{Kaplan2009PRA}, so here 
we only provide a brief overview of the ``$\perp$'' case and outline the 
most distinctive features appearing in this geometry.

Let us start again with describing locsitons in unbounded lattices.
The most general equation for the LF in the ``$\perp$'' case is 
obtained from Eq.~(\ref{eq-LF}) by setting 
$\vect{E}_\mathrm{in} (\vect{r}) \parallel \vect{E}_\mathrm{L} (\vect{r}) 
\perp \vect{u}$:
\begin{equation}
\vect{E}_\mathrm{L} (\vect{r}) = \vect{E}_\mathrm{in} (\vect{r}) 
+ \frac{Q}{4} \sum^{\vect{r}' \ne \vect{r}}_\mathrm{lattice}
\frac{l_a^3 }{ |\vect{r}' - \vect{r}|^3 }
\vect{E}_\mathrm{L} (\vect{r}').
\label{eq-LF-perp}
\end{equation}
As in the ``$\parallel$'' case, we will assume here that the incident optical 
field is uniform, 
$\vect{E}_\mathrm{in} (\vect{r}) \equiv \vect{E}_\mathrm{in}$, which will help 
us to build a clearer understanding of the locsiton behavior in 2D lattices.
For an equilateral triangular lattice, Eq.~(\ref{eq-LF-perp}) can be 
simplified using the NNA as
\begin{equation}
\vect{E}_\mathrm{L} (\vect{r}) = \vect{E}_\mathrm{in}
+ \frac{Q}{4} \sum_{\vect{u}_\mathrm{K}}
    \vect{E}_\mathrm{L} (\vect{r} + l_a\vect{u}_\mathrm{K}).
\label{eq-LF-NNA-perp}
\end{equation}
Equation (\ref{eq-LF-NNA-perp}) has a uniform, Lorentz, solution, which is 
given by Eq.~(\ref{eq-Lorentz}) with 
$\delta_\mathrm{LL} = \delta_\mathrm{LL}^\perp$ where
\begin{equation}
\delta_\mathrm{LL}^\perp = \frac{3}{2} Q_a,
\label{eq-dLL-perp}
\end{equation}
which is three times the NNA value for $\delta_\mathrm{LL}$ in a 1D array of 
atoms if $\vect{E}_\mathrm{in}$ is perpendicular to the array (and the 
dipoles are aligned ``side-to-side'') \cite{Kaplan2009PRA}.
This is a consequence of each atom having now 6 instead of 2 neighbors.

To obtain the corresponding dispersion relation $\delta(\vect{q})$ for 
``$\perp$'' locsitons in the lattice, we substitute the LF 
$\vect{E}_\mathrm{L} (\vect{r})$ in Eq.~(\ref{eq-LF-NNA-perp}) as a sum of the
Lorentz solution given by Eqs.~(\ref{eq-Lorentz}) and~(\ref{eq-dLL-perp}) and 
2D plane-wave excitations (\ref{eq-DeltaE}).
This dispersion relation can be written as
\begin{equation}
D_{2\perp}^\mathrm{NNA}(\vect{q}) 
\equiv \frac{1}{3} \sum_{n=0}^2 \cos[q \cos(\theta_n - \psi)]
= \frac{\delta + i}{\delta_\mathrm{LL}^\perp},
\label{eq-disp-NNA-perp}
\end{equation}
where, as in Sec.~\ref{subsec-BZ}, $\theta_n = \theta_0 + n\pi/3$.
In the long-wavelength limit, 
\begin{equation}
D_{2\perp}^\mathrm{NNA}(\vect{q}) \approx 1 - \frac{1}{4} q^2 + O(q^4),
\label{eq-disp-perp-small-q}
\end{equation}
so that the uniform, Lorentz, solution for the LF is reached at 
$\delta = \delta_\mathrm{LL}^\perp$, where
$D_{2\perp}^\mathrm{NNA}(\vect{q} = \vect{0}) = 1$.
The second term in the r.\,h.\,s.\ of Eq.~(\ref{eq-disp-perp-small-q}) is
independent of the orientation of $\vect{q}$, which means that no anisotropy 
caused by the lattice structure is present in the long-wavelength limit.
We may thus conclude that, compared to the ``$\parallel$'' case, 
the locsitons in the ``$\perp$'' configuration are more reminiscent of the 
locsitons in 1D arrays of resonant atoms considered in 
\cite{Kaplan2008PRL,Kaplan2009PRA}.
There is still no complete analogy here, as, e.\,g., the second term in 
the r.\,h.\,s.\ of Eq.~(\ref{eq-disp-perp-small-q}) differs by a factor 
of $1/2$ from the 1D result \cite{Kaplan2009PRA}.
Moreover, dispersion relation (\ref{eq-disp-NNA-perp}) does become anisotropic 
for larger $q$, closer to the boundaries of the first Brillouin zone.
This anisotropy, however, is by far less pronounced than that in the 
``$\parallel$'' case.

It is instructive to also obtain the dispersion relation in the NRA.
By replacing the summation in Eq.~(\ref{eq-disp-NNA-perp}) with an integration 
over the ``near ring'', following the procedure outlined in 
Sec.~\ref{subsec-NRA}, we get
\begin{equation}
1 - \frac{3Q}{2\pi} \int_0^\pi \cos[q\cos(\theta - \psi)]\, d\theta = 0.
\label{eq-disp-NRA-perp-1}
\end{equation}
The resulting dispersion relation turns out to be independent of the 
orientation of $\vect{q}$:
\begin{equation}
D_{2\perp}^\mathrm{NRA}(\vect{q}) \equiv J_0(q) 
= \frac{\delta + i}{\delta_\mathrm{LL}^\perp},
\label{eq-disp-NRA-perp-2}
\end{equation}
which is not surprising given the NRA applicability in the long-wavelength 
limit.

While it might be somewhat harder to create a uniform incident field polarized 
normally to a 2D lattice, the resulting locsitons could be much easier to 
control because of the small anisotropy of the interatomic interactions in the 
``$\perp$'' geometry, compared to the ``$\parallel$''  geometry.
For example, defects in a 2D lattice can support \emph{localized} locsitons, 
not unlike the \emph{evanescent} 1D locsitons discussed in 
\cite{Kaplan2009PRA}.
Compared to the complex locsiton patterns emerging in the ``$\parallel$'' 
geometry [cf.\ Fig.~\ref{fig-vortices}] these localized locsitons are more 
likely to form well-organized strata-like patterns in the ``$\perp$'' 
geometry.

\begin{figure}
\includegraphics[width=3.25in]{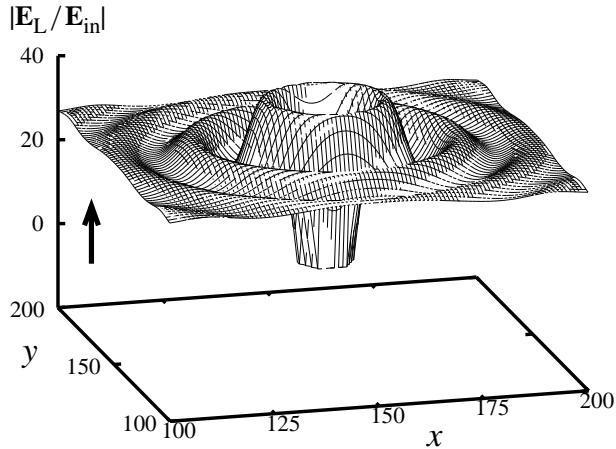}
\caption{%
Spatial strata of the normalized local field around a 15-point-wide hole 
in a 2D triangular lattice at $\delta = 100$ and $\delta_\mathrm{LL} = 103.5$.
The direction of the incident field is shown with a big arrow.}
\label{fig-hole}
\end{figure}

Fig.~\ref{fig-hole} shows concentric dipole strata that are formed around a 
circular hole made by removing a few tens of atoms from a triangular lattice.
The locsiton ``attached'' to the defect ``decays'' as the distance to the hole 
boundary increases, which is mostly a ``diffraction'' effect, although some 
contribution from the imaginary part of $\vect{q}$ (like in evanescent 1D
locsitons) is also present.
In performing the numerical simulations for the plot, we made sure that the 
locsitons attached to the outside boundaries of the lattice patch (lying far
outside the plotted region) do not interfere with the locsiton localized at 
the defect.

\section{Conclusions}
\label{sec-conclusions}

In this paper we presented a detailed study of locsitons in 
nanoscale 2D lattices of resonant atoms with strong dipole interaction.
These locsitons (i.\,e., excitations of the local field) and various 
associated effects were originally predicted in our recent publication 
\cite{Kaplan2008PRL}.
Here, we have built analytic models for locsitons in infinite 2D triangular
lattices of atoms, based either on the nearest-neighbor approximation or
on the simpler near-ring approximation.

We have shown that the ``in-plane'' polarization geometry, 
where the incident laser field lies in the lattice plane,
enables locsitons with the most unusual and diverse 
properties, as compared to the 1D case described in detail in 
\cite{Kaplan2009PRA}.
In particular, the dispersion relations for the locsitons with an in-plane polarization 
are highly anisotropic with respect to the orientation of the locsiton 
polarization relative to its wave vector in the lattice plane, because
of the highly anisotropic nature of the dipole-dipole interaction.
We further demonstrated a method to design a finite 2D lattice, such that 
distinct vector locsiton patterns are formed at a certain laser frequency, 
the patterns containing multiple vortices in the local field distribution.

We have also considered a remarkable effect of a cancellation of the resonant 
local field suppression \cite{Kaplan2008PRL}, 
which consists in the local field being able to penetrate certain 
``magic shapes'' made of resonant 
atoms, despite the nearly-universal tendency 
of the local field to be ``pushed out'' 
of the lattice at the exact atomic resonance.
In particular, we provided more detail on the local field distribution in the 
simplest ``magic shape'' that can be cut out of a triangular lattice---a 
six-point star with an atom at the center.

In the case where the incident field is polarized normally to the lattice,
we found that locsitons bear more analogy to locsitons in 1D arrays of atoms, 
compared to the case of an in-plane polarization.
Finally, we illustrated the role of lattice defects in supporting localized 
locsitons.

While this paper does not elaborate on nonlinear effects involving 2D locsitons,
to be addressed in our future publications, we note that, 
similarly to the 1D case \cite{Kaplan2008PRL,Kaplan2009PRA},
our numerical simulations have shown optical 
bistability and hysteresis, which may be especially 
important for potential applications of 2D locsitons in designing 
all-dielectric nanoscale logic elements, devices for signal processing, etc.

\acknowledgments

This work is supported by US AFOSR.


\end{document}